# Multi-State Pair-Density Functional Theory

Jie J. Bao,[&] Chen Zhou,[&] Zoltan Varga, Siriluk Kanchanakungwankul, Laura Gagliardi,* and Donald G. Truhlar*

*Department of Chemistry, Chemical Theory Center, and Minnesota Supercomputing Institute, 207 Pleasant Street SE, University of Minnesota, Minneapolis, MN 55455-0431, USA*

**Abstract.** Multiconfiguration pair-density functional theory (MC-PDFT) has previously been applied successfully to carry out ground-state and excited-state calculations. However, because it includes no interaction between electronic states, MC-PDFT calculations in which each state's PDFT energy is calculated separately can give an unphysical double crossing of potential energy surfaces (PESs) in a region near a conical intersection. We have recently proposed state-interaction pair-density functional theory (SI-PDFT) to treat nearly degenerate states by creating a set of intermediate states with state interaction; although this method is successful, it is inconvenient because two SCF calculations and two sets of orbitals are required and because it puts the ground state on an unequal footing with the excited states. Here we propose two new methods, called extended-multi-state-PDFT (XMS-PDFT) and variational-multi-state-PDFT (VMS-PDFT), that generate the intermediate states in a balanced way with a single set of orbitals. The former uses the intermediate states proposed by Granovsky for extended multiconfiguration quasidegenerate perturbation theory (XMC-QDPT); the latter obtains the intermediate states by maximizing the sum of the MC-PDFT energies for the intermediate states. We also propose a Fourier series expansion to make the variational optimizations of the VMS-PDFT method convenient, and we implement this method (FMS-PDFT) both for conventional configuration-interaction solvers and for density-matrix-renormalization-group solvers. The new methods are tested for eight systems exhibiting avoided crossings among two to six states. The FMS-PDFT method is successful for all eight test cases studied in the paper, and XMS-PDFT is successful for all of them except the mixed-valence case. Since both XMS-PDFT and VMS-PDFT are less expensive than XMS-CASPT2, they will allow well-correlated calculations on much larger systems for which perturbation theory is unaffordable.

# 1. Introduction

Kohn-Sham density functional theory (KS-DFT)[1] has been successful in treating many chemical problems, but it is less accurate for treating inherently multiconfigurational electronic states – which are called strongly correlated states – than for treating states well represented by a single Slater determinant – which are called weakly correlated.[2] Strong correlation usually arises from near degeneracy of two or more states, and excited electronic states are usually strongly correlated, and often they are strongly interacting with other states. Thus, the accurate treatment of strongly correlated states is necessary for spectroscopy and photochemistry.[3,4] Furthermore, the accurate treatment of strongly correlated sets of states is also required to properly describe magnetic effects.[5,6]

Although KS-DFT has lower accuracy for strongly correlated states than for weakly correlated ones, for large molecules it is much less expensive than wave function theory (WFT) methods of comparable accuracy. We have proposed multiconfiguration pair-density functional theory (MC-PDFT) as a method that builds on a multiconfigurational self-consistent-field (MCSCF) reference wave function and is more innately suitable for strongly correlated systems than KS-DFT; MC-PDFT also has the advantage of being computationally less expensive compared with WFT methods in terms of computer time and memory with comparably accurate treatments of correlation energy.[7,8] We refer the reader to a recent review article[9] that compares PDFT to other ways to combine wave function methods and density functional methods for excited-state calculations.

When states are nearly degenerate and have the same symmetry, they interact strongly with each other, and they should be treated by a method that gives the correct topography[4] of adiabatic potential energy surfaces (PESs) at conical intersections; such methods are called multi-state (MS) methods. For example, in WFT, multireference Møller-Plesset perturbation theory[10] is a state-specific method because it calculates the final approximation to the energy of each state separately, whereas multiconfiguration quasidegenerate perturbation theory (MC-QDPT)[11] or extended MC-QDPT (XMC-QDPT)[12] are multi-state methods because the final energies are eigenvalues of the same matrix (hence they interact through the off-diagonal elements of that matrix). Similarly, complete active space perturbation theory (CASPT2)[13] is a state-specific method, and multi-state CASPT2 (MS-CASPT2)[14] and extended MS-CASPT2 (XMS-CASPT2)[15] are multi-state methods.

The original MC-PDFT is a state-specific method. We recently proposed state-interaction PDFT (SI-PDFT) as a multi-state generalization;[16] SI-PDFT yields the correct topography of adiabatic PESs for conical intersections and it has been applied successfully to several problems;[16,17,18] but it is inconvenient because two MCSCF calculations and two sets of orbitals are required, and it puts the ground state on an unequal footing with the excited states, which is sometimes undesirable (for example, for treating magnetic states). In the present paper we present two new multi-state methods that eliminate these drawbacks of SI-PDFT. One is called extended-multi-state-PDFT (XMS-PDFT) because it uses the intermediate basis proposed by Granovsky[12] for XMC-QDPT, and the other is called variational-multi-state-PDFT (VMS-PDFT) because it obtains an intermediate basis by variationally maximizing the sum of MC-PDFT energies for the intermediate states. We approximate the VMS-PDFT method by using a Fourier series expansion; using this method for VMS-PDFT is called the Fourier-multi-state-PDFT (FMS-PDFT). method.

A key aspect of all the above-mentioned MS methods is that they determine a model space spanned by the states to be treated as strongly interacting. Similar to XMC-QDPT or XMS-CASPT2, XMS-PDFT and VMS-PDFT build up a model space that spans the $N$ lowest-energy states optimized in a state-averaged CASSCF (SA-CASSCF) calculation. (Generalizations to incomplete active spaces and smaller model spaces are straightforward but are not considered here.) The model space states are called the intermediate basis and are obtained by unitary transformation from the SA-CASSCF states.

Section 2 explains the two new methods and the Fourier-based approximation of VMS-PDFT. Section 3 specifies computational details for several test systems, including those that were previously studied by SI-PDFT. Section 4 presents applications of the new methods to these test systems and evaluates their performances. Section 5 has concluding remarks.

## 2. Theory

### 2.1 MC-PDFT

The MC-PDFT method may be based on single-state CASSCF (SS-CASSCF) calculations or on SA-CASSCF calculations. In the present article we consider the latter type of calculation, in which case one starts with a reference wave function obtained by performing an SA-CASSCF calculation and given by

$$|\Psi_I\rangle = \sum_i c_i^I |\text{CSF}_i\rangle, \qquad (1)$$

where $i$ is the index of a configuration state function (CSF), and $I$ is the index of a reference state. The MC-PDFT energy for state $I$ is

$$E_I^{\text{MC-PDFT}} = T_e + V_{\text{elec}} + E_{\text{ot}}(\rho_I, \Pi_I), \qquad (2)$$

where the terms are the electronic kinetic energy, classical electrostatic energy (which is the sum of the nuclear-nuclear repulsion, the electron-nuclear attraction energy, and the classical electron-electron repulsion), and the on-top energy computed as a functional of the density $\rho_I$ and the on-top density $\Pi_I$, both computed from $|\Psi_I\rangle$, with the latter given by

$$\Pi_I(\mathbf{r}) = \int \Psi_I^*(\mathbf{r}_1, \mathbf{r}_2, \ldots, \mathbf{r}_{N_e}) \Psi_I(\mathbf{r}_1, \mathbf{r}_2, \ldots, \mathbf{r}_{N_e}) d\mathbf{r}_3 \ldots d\mathbf{r}_{N_e}|_{\mathbf{r}_1=\mathbf{r}_2=\mathbf{r}}. \qquad (3)$$

Equation (2) applies to MC-PDFT calculations staring with either SS-CASSCF or SA-CASSCF. We note that it does not separate the energy into an uncorrelated component, a static correlation component, and a dynamic correlation component. Because the original MC-PDFT method computes the state energies independently, it is a state-specific method in the sense that the final energy of each state is computed separately, even if one starts with SA-CASSCF kinetic energies, densities, and on-top densities.

### 2.2 Multi-state MC-PDFT

To obtain the correct topography of PESs at conical intersections, we have proposed the SI-PDFT method[16] as an MS extension of MC-PDFT. In SI-PDFT, we generate a set of intermediate

4states with the reference SA-CASSCF states and an auxiliary state from a state-specific ground-state CASSCF calculation. The ground intermediate state is obtained by projecting the SS-CASSCF state into the space spanned by the SA-CASSCF states, and the other intermediate states are obtained by performing Schmidt orthogonalization of the excited states obtained by the SA-CASSCF calculation to the ground intermediate state. Then one constructs an effective Hamiltonian in the intermediate state basis and diagonalizes it to get the SI-PDFT energy for each state. This treats the ground and excited states unequally. Moreover, using different orbital sets (i.e., using both the orbitals from the SS-CASSCF calculation and those from the SA-CASSCF calculation) is inconvenient. To avoid these problems, we next propose two new multi-state MC-PDFT methods that use only one set of orbitals.

In general, the intermediate states are obtained by a unitary transformation:

$$|\Phi_I\rangle = \sum_J U_{JI} |\Psi_J\rangle = \sum_{Ji} U_{JI} c_i^J |\text{CSF}_i\rangle, \qquad (4)$$

where $|\Phi_I\rangle$ is an intermediate state, and $|\Psi_J\rangle$ is an SA-CASSCF state. The Hamiltonian is diagonal in the SA-CASSCF states but not in the intermediate basis.

We construct an effective Hamiltonian in the intermediate-state basis with diagonal elements defined as

$$H_{II}^{\text{eff}} = E_I^{\text{MC-PDFT}}, \qquad (5)$$

where $E_I^{\text{MC-PDFT}}$ is the MC-PDFT energy for the intermediate state $|\Phi_I\rangle$. The off-diagonal elements of the effective Hamiltonian are defined as

$$H_{IJ}^{\text{eff}} = \langle \Phi_I | H | \Phi_J \rangle \qquad (6)$$

with $I, J = 1, 2, ..., N$, where $N$ is the number of states in the model space. (In the present work, the number of states in the model space is always the same as the number of states averaged in the SA-CASSCF calculation.) The effective Hamiltonian is then diagonalized to give the multi-state MC-PDFT energies for each adiabatic state.

Following the above scheme, we next introduce two strategies (XMS-PDFT and VMS-PDFT) to generate the matrix $\mathbf{U}$, yielding $\mathbf{U}^X$ and $\mathbf{U}^V$, respectively.

**2.3 XMS-PDFT**

The intermediate basis in XMS-PDFT diagonalizes the effective Hamiltonian suggested by Granovsky for XMC-QDPT in ref. 12, where he stressed that "the effective Hamiltonian should be a function of the subspace spanned by the selected CI vectors, rather than a function of any particular choice of basis in this subspace" and that "the computed energies must be uniquely defined, continuous and smooth functions of the molecular geometry and any other external parameters, with possible exceptions at the manifolds of their accidental degeneracy such as conical intersections". The XMS-CASPT2 method also uses this intermediate basis. We use the XMS-CASPT2 procedure[15] to explain this, and the explanation starts by recalling the procedure in MS-CASPT2.

In MS-CASPT2, the unperturbed Hamiltonian is defined as

$$H_0 = PFP + QFQ, \qquad (7)$$

where

$$P = \sum_I |\Psi_I\rangle\langle\Psi_I|$$

is the projection operator onto the SA-CASSCF state space and

$$Q = 1 - P$$

is the projection operator onto the complementary state space. In MS-CASPT2, the state Fock operator of states is defined as

$$F = \sum_{pq} f_{pq} E_{pq} = \sum_{pq} f_{pq} a_p^\dagger a_q, \qquad (8)$$

where $E_{pq} = a_p^\dagger a_q$ is a single-excitation operator, $a_p^\dagger$ and $a_q$ are creation and annihilation operators on molecular orbitals $p$ and $q$, respectively, and $f_{pq}$ is an element in the orbital Fock matrix

$$f_{pq} = h_{pq} + \sum_{rs} d_{rs}(J_{pq}^{rs} - \tfrac{1}{2} K_{pq}^{rs}), \qquad (9)$$

where $h_{pq}$ contains the electronic kinetic energy and electron-Coulomb interaction, $d_{rs}$ is a state-averaged density matrix element, and $J_{pq}^{rs}$ and $K_{pq}^{rs}$ are two-electron integrals. The matrix elements of the state Fock matrix are defined as

$$F_{IJ} = \langle \Psi_I | F | \Psi_J \rangle = \sum_{pq} \sum_{ij} f_{pq} c_i^I c_j^J \langle \text{CSF}_i | E_{pq} | \text{CSF}_j \rangle. \qquad (10)$$

The state Fock matrix defined in eqn (10) is not necessarily diagonal, because the reference wave functions (i.e., the SA-CASSCF wave functions) are the eigenstates of the Hamiltonian operator, not necessarily the eigenstates of the state Fock operator or the zeroth-order Hamiltonian.

The MS-CAPST2 method neglects the off-diagonal elements of the state Fock matrix, but following the prescription used in the XMS-CASPT2 method, the XMS-PDFT method diagonalizes the state Fock matrix by a transformation matrix $\mathbf{U}^X$:

$$(\mathbf{U}^X)^\dagger \mathbf{F} \mathbf{U}^X = \widetilde{\mathbf{F}}. \qquad (11)$$

The $\mathbf{U}^X$ matrix determined this way then yields the intermediate states defined by

$$|\Phi_I\rangle = \sum_J U_{JI}^X |\Psi_J\rangle, \qquad (12)$$

where $\Phi_I$ is an intermediate state in XMS-PDFT (and also in XMS-CASPT2). With the same transformation, we get a Hamiltonian matrix in the intermediate basis,

$$(\mathbf{U}^X)^\dagger \mathbf{H} \mathbf{U}^X = \widetilde{\mathbf{H}}, \qquad (13)$$

where $\mathbf{H}$ is the Hamiltonian matrix in the basis of the SA-CASSCF reference states, and $\widetilde{\mathbf{H}}$ is in the basis of the intermediate states.

After the intermediate states are obtained, XMS-PDFT defines an effective Hamiltonian in the intermediate basis such that diagonal element $H_{II}^{\text{eff}}$ is the MC-PDFT energy of intermediate state $\Phi_I$, and the off-diagonal element $H_{IJ}^{\text{eff}}$ is $\widetilde{H}_{IJ}$. The XMS-PDFT energies ($E_I^{\text{XMC-PDFT}}$) and eigenvectors are obtained by diagonalizing the effective Hamiltonian matrix.

We notice that the off-diagonal elements in the state Fock matrix are zero for states with different symmetries. This suggests that the XMS-PDFT method is identical to MC-PDFT if all states in the model space belong to different irreps (similarly, XMS-CASPT2 is identical to MS-CASPT2 or even single-state CASPT2 for such a case). This is not a problem, but we have found that the off-diagonal elements in the state Fock matrix are almost zero for many geometries in some mixed-valence[19] systems (see Fig. S3, where sections and figures with the prefix S are in Supporting Information) even when the states have the same symmetry, and we will see that XMS-PDFT does not always give good results for such systems. Next we present the VMS-PDFT method that does not have this problem (but it is more expensive).

**2.4 VMS-PDFT**

The trace of the effective Hamiltonian defined above is given by

$$\text{Tr}(\mathbf{H}^{\text{eff}}) = \sum_I E_I^{\text{MC-PDFT}} \tag{14}$$

Although the wave function contribution to an MC-PDFT energy (i.e., the first two terms of eqn (2)) is unitarily invariant, the on-top energy is not, and therefore the trace in eqn (14) depends on the transformation matrix $\mathbf{U}$. In VMS-PDFT, the transformation matrix $\mathbf{U}^V$ that yields the intermediate basis is chosen so that this trace is maximized. Just as in XMS-PDFT, but using the new intermediate basis, VMS-PDFT then evaluates the energies by diagonalizing an effective Hamiltonian defined such that the diagonal elements are MC-PDFT energies in the intermediate basis and the off-diagonal elements are computed by standard wave function theory in the intermediate basis.

The motivation for using a transformation that maximizes the sum of on-top energies for intermediate states is a physical one, namely that the diagonalization of the effective Hamiltonian can be interpreted as adding extra correlation to the energies of intermediate states, so this correlation energy should not already be present in the diagonal elements.

At present, we do not have an analytic procedure to find the transformation matrix $\mathbf{U}^V$ that completely maximizes eqn (14). Instead, we propose here a numerical way to approximate the maximization in a practical and smooth way by fitting eqn (14) to a Fourier series. We call this implementation the FMS-PDFT method.

We first present FMS-PDFT for a two-state calculation. The unitary transformation between two states $(\Psi_I, \Psi_J)$ can be parameterized as

$$\mathbf{U}_{IJ}^V(\theta_{IJ}) = \begin{pmatrix} \cos\theta_{IJ} & \sin\theta_{IJ} \\ -\sin\theta_{IJ} & \cos\theta_{IJ} \end{pmatrix}, \tag{15}$$

where $\theta_{IJ}$ is the rotation angle between the two states. Applying $\mathbf{U}_{IJ}^V(\theta_{IJ})$ to a pair of states yields

$$(\Psi_I, \Psi_J)\mathbf{U}_{IJ}^V(\theta_{IJ}) = (\Phi_I, \Phi_J). \tag{16}$$

Now consider applying $\mathbf{U}(\theta + \frac{\pi}{2})$ to the two states; this yields

$$(\Psi_I, \Psi_J)\mathbf{U}_{IJ}^V\left(\theta_{IJ} + \frac{\pi}{2}\right) = (-\Phi_J, \Phi_I) \tag{17}$$

Comparing eqns (16) and (17) shows that $\mathbf{U}_{IJ}^V(\theta_{IJ})$ and $\mathbf{U}_{IJ}^V\left(\theta_{IJ} + \frac{\pi}{2}\right)$ generate the same two states but with different ordering, and thus they give the same trace of the effective Hamiltonian matrix. This means that the trace of effective Hamiltonian matrix has a period of $\frac{\pi}{2}$, and therefore the Fourier expansion of the effective Hamiltonian can be written as

$$\text{Tr}(\mathbf{H}^{\text{eff}}) = \frac{a_0}{2} + \sum_{n=1}^{\infty}[a_n \sin(4n\theta_{IJ}) + b_n \cos(4n\theta_{IJ})]. \tag{18}$$

We keep only the terms with $n = 1$ in the sum on the right-hand side of eqn (18); then the equation can be parameterized as

$$\text{Tr}(\mathbf{H}^{\text{eff}}) = A + B \sin(4\theta_{IJ}) + C \cos(4\theta_{IJ}), \tag{19}$$

and the unknown parameters *A*, *B*, and *C* can be obtained by a three-point fitting. In this paper, three values (0º, 30º, and 60º) for $\theta_{IJ}$ are applied to determine these three parameters for each single-point energy calculation. (We use these same three angles in all cases.) Then rotation angle is taken as the one that maximizes eqn (19).

For FMS-PDFT calculations with *N* states (where *N* is greater than 2), we write $\mathbf{U}^V$ as a product of transformation matrices

$$\mathbf{U}^V = \mathbf{U}_{12}\mathbf{U}_{23}\cdots\mathbf{U}_{I(I+1)}\cdots\mathbf{U}_{(N-1)N}, \quad (20)$$

where $\mathbf{U}_{I(I+1)}(\theta_{I,I+1})$ is a unitary matrix that rotates states *I* and (*I*+1) as in eqn (15). When each $\mathbf{U}_{I(I+1)}$ is applied, we have

$$\mathbf{\Phi}^{(I)} = \mathbf{\Phi}^{(I-1)}\mathbf{U}_{I(I+1)}, \quad (21)$$

where *I* ranges from 1 to *N*-1, and $\mathbf{\Phi}^{(I)}$ denotes the *N* intermediate states after transformation of $\mathbf{U}_{12}\mathbf{U}_{23}\cdots\mathbf{U}_{I(I+1)}$, and $\mathbf{\Phi}^{(0)}$ denotes the initial states, which are SA-CASSCF states in this paper. Thus, eqn (19) can also be applied to fitting the trace of effective Hamiltonian for each unitary transformation $\mathbf{U}_{I(I+1)}$.

Notice that neither do we include all the *N*(*N*−1)/2 transformation matrices $\mathbf{U}_{IJ}$ (*I* < *J*) nor do we transform the states iteratively to reach the absolute maximum trace. Since strong couplings mostly occur between adjacent states, it is a reasonable and economical practice to consider the only (*N*-1) unitary transformations $\mathbf{U}_{I(I+1)}$ as presented above. Furthermore, we will see below that the results are already good with a single pass as in eqn (20). Furthermore, stopping with a single pass gives smoother results than one would obtain if one used a convergence criterion that lead to different numbers of iterations at different geometries.

The FMS-PDFT method is also implemented for wave functions optimized with the density matrix renormalization group[20,21,22,23,24,25] (DMRG) approach; the combination of state-specific PDFT and DMRG was introduced previously[26,27] and is here extended to an MS treatment. The FMS-PDFT/DMRG method is extension of the FMS-PDFT method described above except that it is based on an SA-DMRG[28] calculation instead of an SA-CASSCF starting point, and this requires a change in implementation since the DMRG wave function is not explicitly expanded in a CSF basis. Therefore we do not obtain intermediate states with eqn (4); instead, the MC-PDFT energies for intermediate states are calculated with the transformed one-body and two-body density matrices,

$$\widetilde{D}^{II}_{pq} = \sum_{JK} U_{JI}U_{KI}D^{JK}_{pq} = \sum_{JK} U_{JI}U_{KI}\langle\Psi_J|E_{pq}|\Psi_K\rangle, \quad (22)$$

$$\widetilde{D}^{II}_{pqrs} = \sum_{JK} U_{JI}U_{KI}D^{JK}_{pqrs} = \sum_{JK} U_{JI}U_{KI}\langle\Psi_J|E_{pq}E_{rs} - \delta_{qr}E_{ps}|\Psi_K\rangle, \quad (23)$$

where $\mathbf{D}^{JK}$ and $\mathbf{d}^{JK}$ are one-body and two-body transition density matrices between the reference states *J* and *K*, and $\widetilde{\mathbf{D}}^{II}$ and $\widetilde{\mathbf{d}}^{II}$ are one-body and two-body density matrices for the intermediate state *I*.

## 3. Computational details

The calculations are performed in *OpenMolcas* v18.09, tag 548-g19e2926-dirty,[29] with codes modified to perform XMS-PDFT and FMS-PDFT calculations. The DMRG calculations are performed with the *QCMaquis* software suite[25,30,31,32] in *OpenMolcas* v18.11, tag 17-g792ff65-dirty, which is modified to perform FMS-PDFT/DMRG calculations.

In XMS-CAPST2 calculations, an ionization-potential-electron-affinity (IPEA) shift[33] of 0.25 a.u. is used. In the FMS-PDFT/DMRG calculations for phenol, we used the same active space as we used for regular FMS-PDFT calculation. The bond dimension ($M$) is set to 500. In the PDFT calculations, we used the translated PBE (tPBE) on-top functional.

Table 1 presents the wave function symmetry, basis set, number of averaged states, number of active electrons, and identities of active MOs for each system studied. The internal coordinates that are scanned for each system are shown in Table 2. The geometries are available in Section S1.

For the HNCO calculations with bond length $r$(NC) = 2.0 – 2.5 Å and torsion $\tau$(HNCO) = 150° (discussed in section 4.3), we carried out the VMS-PDFT calculations using a numerical maximization procedure instead of the Fourier series algorithm because the 3-point fitting in FMS-PDFT fails for that limited region due to the trace of the effective Hamiltonian changing slowly with respect to the rotation angle, so that keeping only the terms with $n = 1$ in the sum on the right hand side of eqn (18) is inadequate. In all other cases the Fourier series method proved adequate.

Table 1. Systems studied, symmetry enforced on the wave function (Sym), basis set, number of states in the SA calculation ($N_{states}$), number of active electrons ($n$) and active molecular orbitals (active MOs)

| System | Sym | Basis set | $N_{states}$ | $n$ | active MOs |
|---|---|---|---|---|---|
| LiF | $C_1$ | jun-cc-pVQZ[34,35] | 2 | 8 | $2p_z$ of F, 2s of Li |
| LiH | $C_{2v}$ | aug-cc-pVQZ[34] | 4 | 2 | 2s, $2p_z$, 3s, $3p_z$ of Li, 1s of H |
| HNCO | $C_1$ | cc-pVDZ[34] | 2 | 16 | Valence shell (2s and 2p of C, N, and O atoms and 1s of H atom) |
| CH$_3$NH$_2$ | $C_1$ | 6-31++G(d,p)[36,37] | 2 | 6 | 2 σ, 1 σ*, $2p_z$, 3s, and $3p_z$ of N |
| C$_6$H$_5$OH | $C_1$ | jul-cc-pVDZ[34,35] | 2 or 3 | 12 | 3 π, 3 π*, $\sigma_{OH}$, $\sigma^*_{OH}$, $\sigma_{CO}$, $\sigma^*_{CO}$ and $p_z$ of O |
| O + O$_2$ ($^3A'$) | $C_s$ | cc-pVTZ[34] | 6 | 12 | 9 2p orbitals |
| O$_3$ ($^3A'$) | $C_s$ | cc-pVTZ[34] | 6 | 12 | 9 2p orbitals |
| Spiro | $C_{2v}$ | 6-31G(d)[38] | 2 | 11 | See Ref. 17 |

Table 2. Systems studied and the internal coordinates scanned for potential energy curves

| System | Internal coordinates scanned |
|---|---|
| LiF | $r$(LiF) = [1.0 – 9.0] Å |
| LiH | $r$(LiH) = [1.0 – 12.0] Å |
| HNCO | $r$(NC) = [1.25 – 3.00] Å and $\tau$(HNCO) = [180 – 130]° |
| CH$_3$NH$_2$ | $r$(NH) = [0.8 – 3.6] Å and $\tau$(H6-C4-N1-H3) = 0, 90, 95, or 100° |
| C$_6$H$_5$OH | $r$(OH) = [0.5 – 3.0 ] Å and $\tau$(C-C-O-H) = 1 or 10° |
| O + O$_2$ ($^3A'$) | $r$(O1O3) = [1.0 – 2.5] Å |
| O$_3$ ($^3A'$) | $\alpha$(O2O1O3) = [60 – 180]° |
| Spiro | See Section 4.8 |

## 4. Results and discussion

### 4.1 Lithium fluoride (LiF)

Lithium fluoride has an avoided crossing of the ground state and first excited state that has been widely studied.[39,40,41,42,43,44,45,46] The ground state at the equilibrium distance is ionic, corresponding to the $(2p_{z,F})^2(2s_{Li})^0$ configuration. The ground state has an $A_1$ symmetry in the $C_{2v}$ group. This state interacts with another $A_1$ state that corresponds to two neutral ground-state atoms, namely $(2p_{z,F})^1(2s_{Li})^1$. The accurate value of the distance of the avoided crossing is about 7.4 Å.[39] However, theoretical calculations usually underestimate the region by 1.0 Å, with an exception being the calculation in Ref. 41.

The MC-PDFT method gives an unphysical double crossing between 4 Å and 6 Å, associated with a "dip" of the energy curve, as shown in Fig. 1(a) and (c). The XMS-PDFT and FMS-PDFT methods, however, remove the incorrect double crossing and also recover the expected shape of the avoided crossing at a larger distance. Additionally, the two new multi-state PDFT methods preserve the correct asymptotic character of the two states, and they work well for the whole potential energy curve.

Figs. 1(b) and (d) show that the XMS-PDFT and FMS-PDFT results agree with XMS-CASPT2 for overall shapes of the two curves. The minimum separation of the two curves is 0.18 eV at 5.97 Å by XMS-PDFT, 0.15 eV at 5.92 Å by FMS-PDFT, and 0.11 eV at 6.11 Å by XMS-CASPT2. The bond lengths with the minimum energy separation by two methods are significantly shorter than 7.2 Å because the calculations underestimate the electron affinity of F, which is a very hard[47] problem.

To check the fitting accuracy of FMS-PDFT, we compare the trace of the effective Hamiltonian matrix obtained by 3-point fitting to that obtained by non-fitted calculations. In Table 3, we list the rotation angles for various Li-F bond lengths and the corresponding trace of effective Hamiltonian obtained by calculations with and without fitting. The MUE of the trace in the fitted calculation is less than 3 meV, which is much less than the intrinsic error in the method and is adequate for most applications.

Table 3. The rotation angles for various Li-F bond lengths and the difference of trace of the effective Hamiltonian obtained by fitting and by specific calculation at that rotation angle.

| $R_{Li-F}$ (Å) | Rotation angle (deg) | ΔE (eV) |
|---|---|---|
| 0.8 | 26.34 | -0.0016 |
| 1.6 | 5.06 | 0.0078 |
| 2.4 | 15.72 | 0.0008 |
| 3.2 | 28.66 | -0.0026 |
| 4.0 | 38.46 | -0.0015 |
| 4.8 | 13.00 | -0.0102 |
| 5.6 | 4.10 | 0.0033 |
| 6.4 | 1.37 | 0.0020 |
| 7.2 | 0.42 | 0.0007 |
| 8.0 | 0.07 | 0.0001 |
| 10.0 | 0.14 | -0.0003 |

| | MUE | 0.0028 |
|---|---|---|

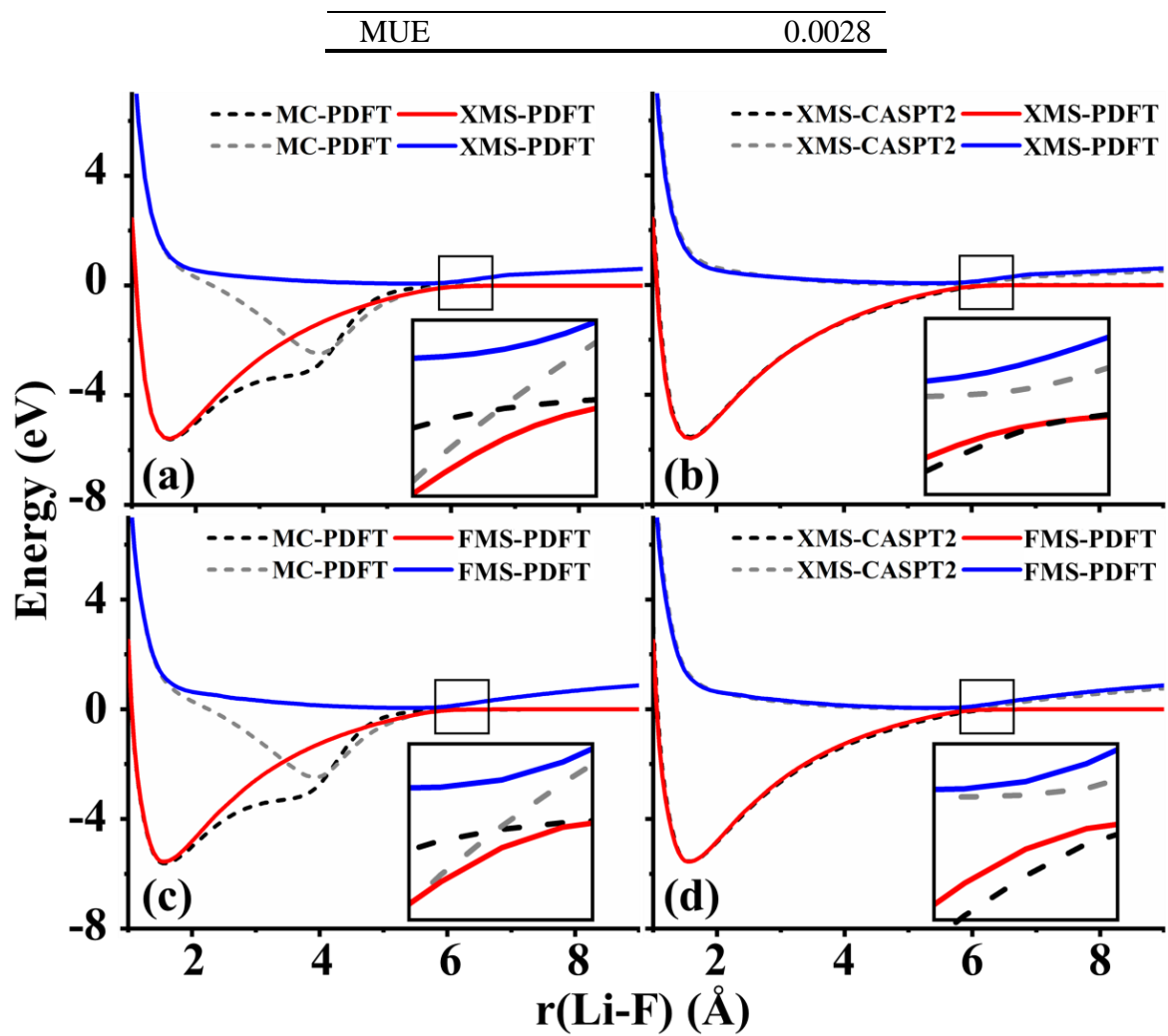

Fig. 1. Comparison of XMS-PDFT, FMS-PDFT, MC-PDFT, and XMS-CASPT2 for the potential energy curves of the two electronic states of LiF. The area near the avoided crossing (indicated by a small box) for each curve is also shown magnified.

### 4.2 Lithium hydride (LiH)In ALL

The ground state of lithium hydride is an ionic state near equilibrium, but this state interacts with three covalent states, corresponding to Li(2s)H(1s), Li(2p$_z$)H(1s), and Li(3s)H(1s) configurations, as the Li-H bond dissociates.

Despite the complexity that the ionic state of LiH crosses with at least three other states as shown in Fig. 2, a similar pattern to LiF is still found for the third and fourth state of LiH beyond 10 Å. The zoomed-in regions in Figs. 2(a) and (c) show the MC-PDFT curves for the third and the fourth states still have a dip and a double crossing, while XMS-PDFT and FMS-PDFT recover the avoided crossing of the two states and also remove the dip.

The first (red) and second (blue) states calculated by XMS-PDFT and FMS-PDFT agree very well with those of XMS-CASPT2. The minimum energy separation between the third and the fourth state is 0.10 eV at 10.66 Å by XMS-PDFT, 0.08 eV at 11.63 Å by FMS-PDFT, and 0.07 eV at 11.28 Å by XMS-CASPT2. The shapes of the XMS-CASPT2 curves match much better with XMS-PDFT and FMS-PDFT than with MC-PDFT, especially for the energy minima of the excited states.

The potential energy curves of XMS-CASPT2 and FMS-PDFT overlap very well, demonstrating the superiority of physically motivated FMS-PDFT, and also showing that applying only ($N$-1) rotations to an $N$-state calculation without iteration is sufficient for FMS-PDFT. Note that there is a bump around 3 Å for the FMS-PDFT potential energy curve. A similar but more indistinct bump can also be found on the XMS-PDFT potential energy curve. The bumps are a result of the interaction of the 4$^{th}$ state and the next higher one, which is not included in calculation. This is a problem not just with the methods presented here but with potential curves calculated by any method based on SA-CASSCF; the highest included state usually has an avoided crossing with the first unincluded state, and this causes some nonsmoothness in potential curves.

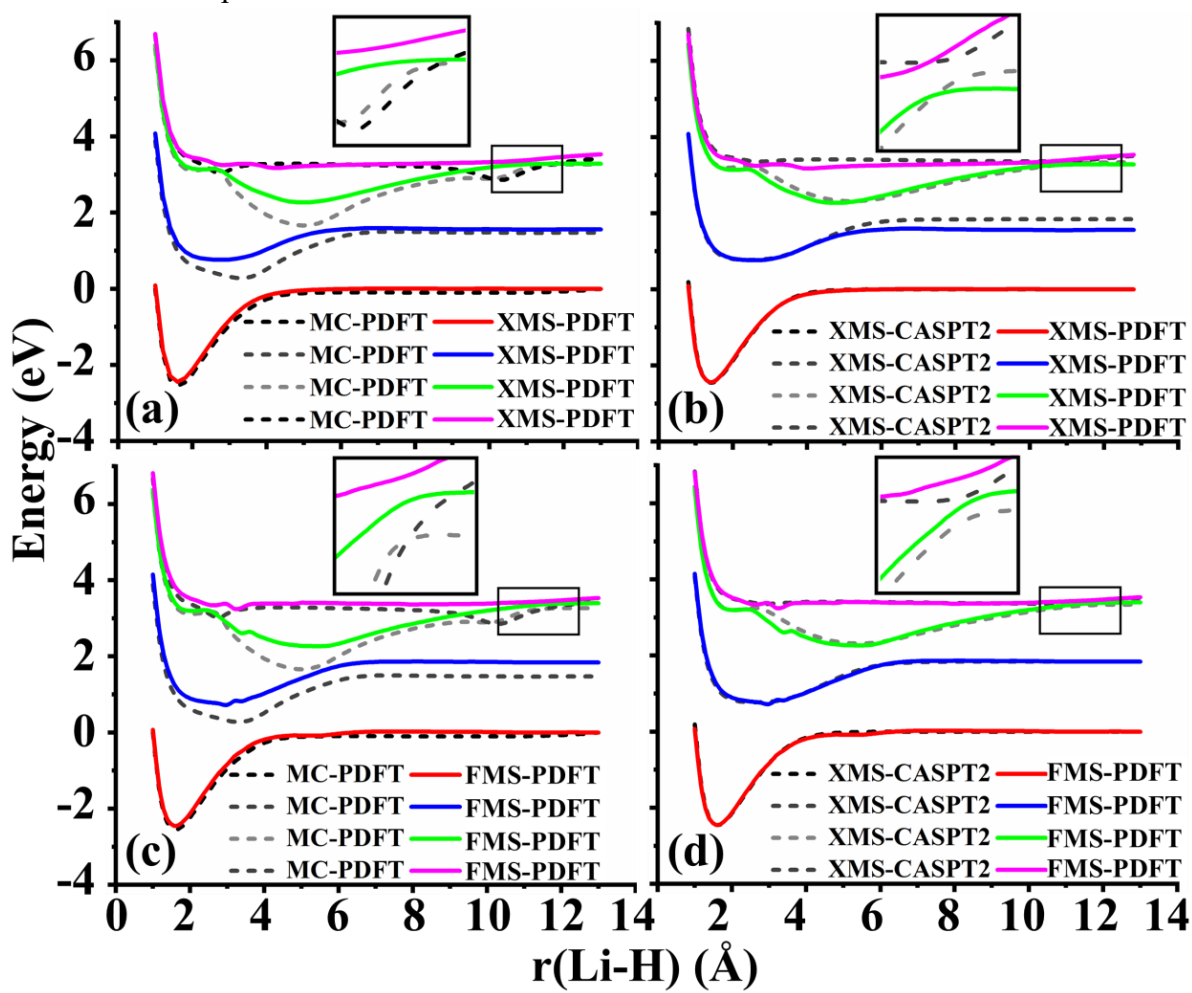

Fig. 2. Comparison of XMS-PDFT, FMS-PDFT, MC-PDFT, and XMS-CASPT2 for the potential energy curves of LiH. The zoomed-in area near the avoided crossing is shown for the ionic state and the highest covalent state calculated.

### 4.3 Isocyanic acid (HNCO)

We next turn to avoided-crossing regions in polyatomics, and we remind the reader that avoided crossings along a polyatomic path are a signal that one is close to a conical intersection.[48]

For planar HNCO, the first two singlet states have $A'$ and $A''$ symmetry in the $C_s$ point group. Along the dissociation path of the NC bond, these two states cross each other. However, if the molecule becomes nonplanar, the crossing becomes avoided (since their coupling becomes symmetry-allowed), and the states switch their character after the avoided crossing. Based on previous work in our group,[49] we fixed two bond lengths, $r(HN) = 1.0584$ Å (2.0 $a_0$) and $r(CO) = 1.1906$ Å (2.5 $a_0$) and two bond angles, $\alpha(HNC) = 110°$, and $\beta(NCO) = 100°$ and then varied the $r(NC)$ bond length from 1.25 to 3.00 Å and the $\tau(HNCO)$ torsion angle from 180 to 130°. We present the curves when the torsion angle is 150° and 175° in Fig. 3 and the curves at other torsion angles are shown in Fig. S1.

Fig. 3 shows that the shapes of the potential curves predicted by XMS-PDFT are similar to those obtained with XMS-CASPT2, although XMS-PDFT predicts a slightly wider energy separation (about 0.17-0.22 eV) of the two states in the region of the equilibrium well and around the region of the energy barrier close to the planar geometry. Fig. S1 shows that further from the planar geometry , with the $\tau(HNCO)$ torsion angle less than 140°, the wider energy separation of XMS-PDFT still holds for the region of the energy barrier, but in the region of the equilibrium well the energy separation of XMS-PDFT becomes slightly narrower than that of XMS-CASPT2. Despite these minor differences between the curves of the two methods, both methods are successful in showing the avoided crossing. Close to the equilibrium geometry, XMS-PDFT places the avoided crossing at about 0.1 Å shorter N-C distance than that of XMS-CASPT2. Since the two states by XMS-PDFT have a slightly wider separation, the avoidance of the two states are more obvious than the avoidance by XMS-CASPT2. Further from the planar geometry, the crossing point of the two states moves to a shorter N-C distance, and we find that the two states are still avoiding each other smoothly.

The torsion angle of 150° is chosen to test the performance of the FMS-PDFT method. A similar avoided crossing around 1.7 Å is observed for both FMS-PDFT and XMS-CASPT2. Although the separation of the two states in the region of 2.0 - 2.5 Å is still wider for VMS-PDFT compared with XMS-CASPT2, VMS-PDFT is shown to be well-behaved, even though we used a combination of FMS-PDFT and numerical VMX-PDFT in this case, as discussed in Section 3.

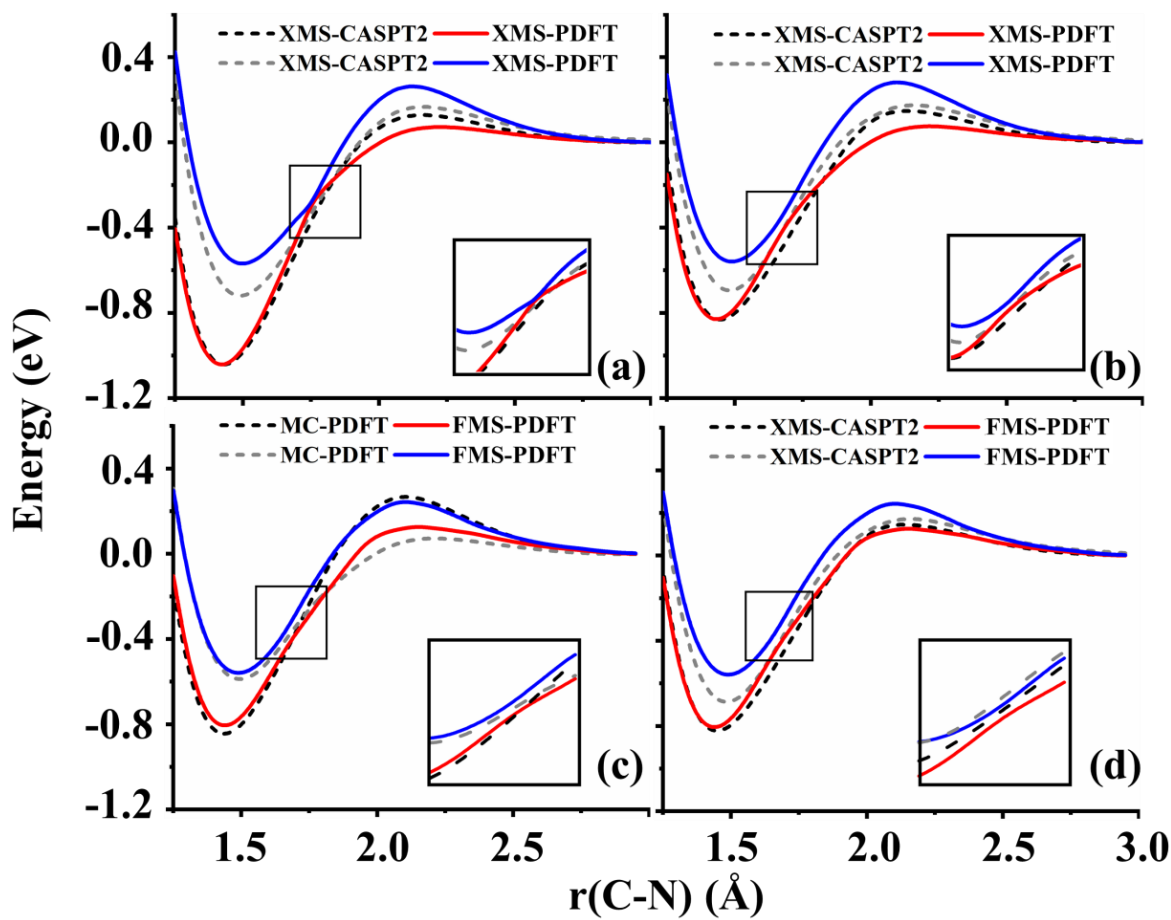

Fig. 3. (a), (b) Comparison of XMS-PDFT with XMS-CASPT2 for the two lowest potential energy curves along the CN bond dissociation of HNCO with the τ(HNCO) torsion angle at 175° and 150° respectively. (c), (d) Comparison of FMS-PDFT with MC-PDFT and XMS-CASPT2 for the two lowest potential energy curves of along the CN bond dissociation with the τ(HNCO) torsion angle at 150°. The curves from 2.0 to 2.5 Å are calculated with numerically optimized VMS-PDFT instead of FMS-PDFT.

**4.4 Methylamine ($CH_3NH_2$)**

The potential energy surfaces, dynamics, and spectroscopy of $CH_3NH_2$ have been widely studied both experimentally and theoretically.[18,50,51,52,53,54,55,56,57,58,59,60,61,62,63] Due to the involvement of the conical intersection region in the photodissociation of methylamine, computational methods should be chosen carefully to correctly describe the strong couplings between the electronic states. The ground and first-excited singlet state were studied along four N-H bond dissociation potential energy curves with XMS-PDFT and FMS-PDFT. These four paths correspond to the N-H bond fissions with conformations shown in Fig. 1 in Ref. 18; these conformations are denoted as eclipsed-H3, staggered, 95º and, 100º, respectively. FMS-PDFT was tested only for the staggered conformation, but MC-PDFT and XMS-PDFT were tested for all four.

    The calculated potential energy curves along the four paths are plotted in Fig. 4. The potential energy curves calculated by both new methods show correct topographies for avoided

crossings near the conical intersection seam, both globally and in the zoomed-in regions. However, the distance at which the minimum energy separation occurs is predicted to be shorter by MC-PDFT and XMS-PDFT than by XMS-CASPT2. The N-H bond distances at the minimum-energy separation and the corresponding energy separations are listed in Table 4. For all the four paths, the PDFT bond distances at the avoided crossing are about 0.056 Å longer than predicted by XMS-CASPT2. However, Table 4 also shows that the minimum-energy separations all agree within 0.08 eV.

Although MC-PDFT does not diagonalize an effective Hamiltonian matrix in the last step, we note that MC-PDFT still gives correct topographies of PESs for the tested four paths of methylamine. Generally speaking, though, MC-PDFT cannot be trusted for regions near conical intersections.

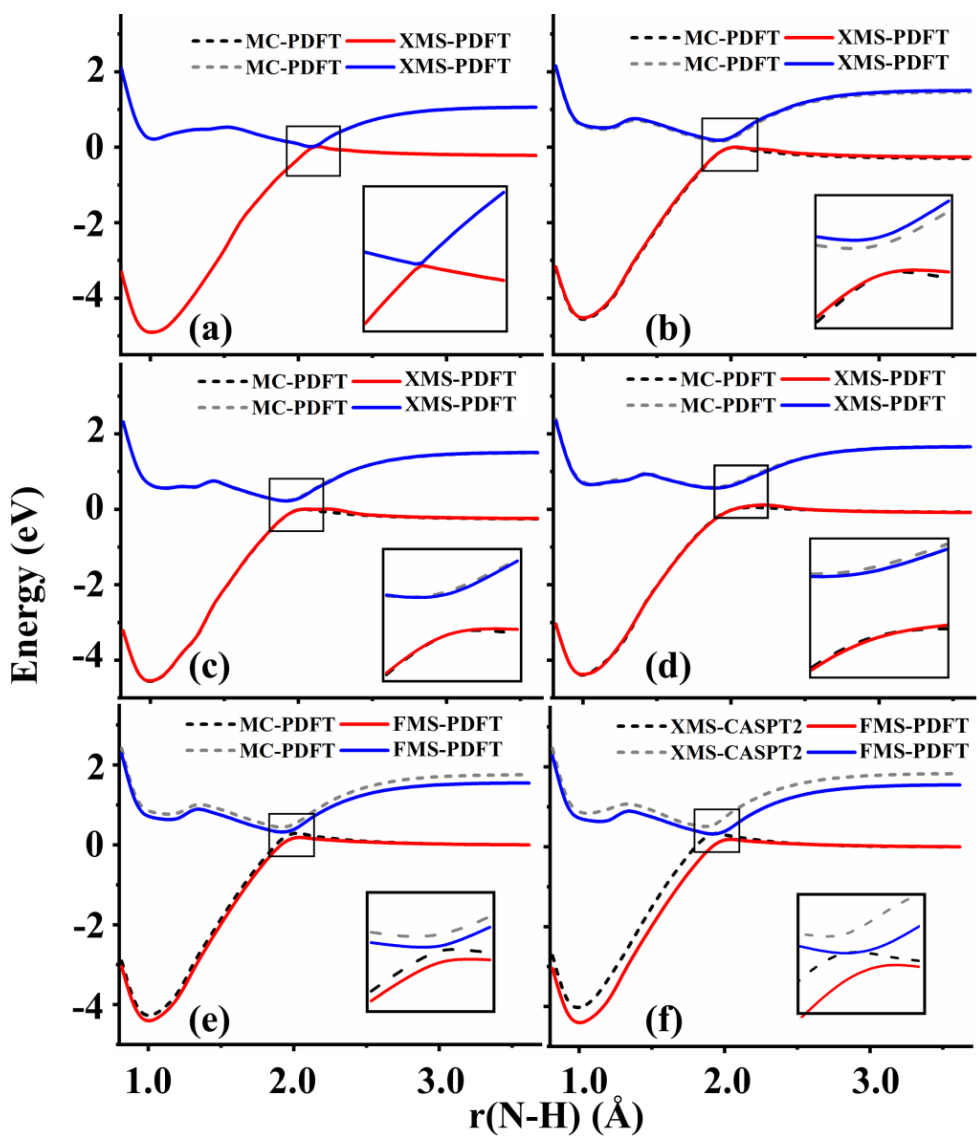

Fig. 4. potential energy curves for methylamine with the four dissociation paths of (a) eclipsed-

H3, (b) staggered, (c) 95º and (d) 100º conformations calculated by XMS-PDFT compared with MC-PDFT. potential energy curves for methylamine with the staggered conformation calculated by FMS-PDFT compared with (e) MC-PDFT and (f) XMS-CASPT2.

Table 4. The N-H bond length (Å) and the energy separations (eV) at the avoided crossing point for the four N-H fission paths.

| Method | $R_{N-H}$ (Å) | Energy separations (eV) |
|---|---|---|
| | Elipsed-H3 | |
| XMS-CASPT2 | 2.01 | 0.007 |
| MC-PDFT | 2.07 | 0.010 |
| XMS-PDFT | 2.07 | 0.010 |
| | Staggered | |
| XMS-CASPT2 | 1.91 | 0.20 |
| MC-PDFT | 1.97 | 0.18 |
| XMS-PDFT | 1.97 | 0.23 |
| FMS-PDFT | 1.97 | 0.18 |
| | 95º | |
| XMS-CASPT2 | 1.92 | 0.33 |
| MC-PDFT | 1.98 | 0.30 |
| XMS-PDFT | 1.98 | 0.28 |
| | 100º | |
| XMS-CASPT2 | 1.95 | 0.73 |
| MC-PDFT | 2.00 | 0.68 |
| XMS-PDFT | 2.02 | 0.65 |

### 4.5 Phenol ($C_6H_5OH$)

The O-H bond dissociation in phenol has been well studied in the past and it can be used as a model system for testing whether a method gives a proper description of potential energy curves for photodissociation. We tested MC-PDFT, XMS-PDFT, FMS-PDFT, and FMS-PDFT/DMRG for the O-H dissociation in phenol with the H-O-C-C dihedral angle being 1° (nearly planar) or 10°.

Fig. 5(a) shows that the MC-PDFT potential energy curves are qualitatively wrong at both angles, with a double crossing when the dihedral angle is 1° and a lack of avoidance at 10° of the torsion. The XMS-PDFT method successfully produces avoided crossings near 2.2 Å for both torsion angles with minimum energy separations of 0.04 and 0.28 eV for 1° and 10°, respectively. The corresponding O-H distances are 2.21 and 2.15 Å.

In regions that are far away from the avoided crossings for each dihedral angle, the XMS-PDFT curves agree very well with the MC-PDFT ones. However, we noticed that XMS-PDFT for this molecule presents a noticeable "bump" after the avoided crossing. This is apparently because the geometry dependence of the off-diagonal elements of the effective Hamiltonian matrix is not consistent enough with the geometry dependence of the diagonal elements. However, the bumps are no greater than 0.07 eV, corresponding to 1.6 kcal/mol, which is usually accurate enough for treating electronically excited states.

The FMS-PDFT method is tested for the O-H dissociation with the H-O-C-C dihedral angle at 10°. Fig. 5(c) shows that FMS-PDFT also succeeds in removing the unphysical double crossing of MC-PDFT. Similar to the issue discussed in regard to the LiH test, the bump near 1.3 Å for both of the states again results from an interaction between the highest included state and the lowest unincluded state. This analysis is confirmed by Fig. 6, which shows that including the third state in the model space replaces the bump by an avoided crossing. However, a new bump now occurs near 1.5 Å due to the interaction between the third and the uninvolved fourth states. The bump could be removed by involving more states in the SA-CASSCF calculation, but the bump due to the interaction between the highest involved state in the SA-CASSCF calculation and higher states not included in the SA calculation is inevitable (although occasionally one is lucky enough that this only occurs at such a high energy as to be insignificant for practical purposes). This is another case that shows it is sufficient in FMS-PDFT to consider only ($N$-1) rotations between adjacent states in an $N$-state calculation.

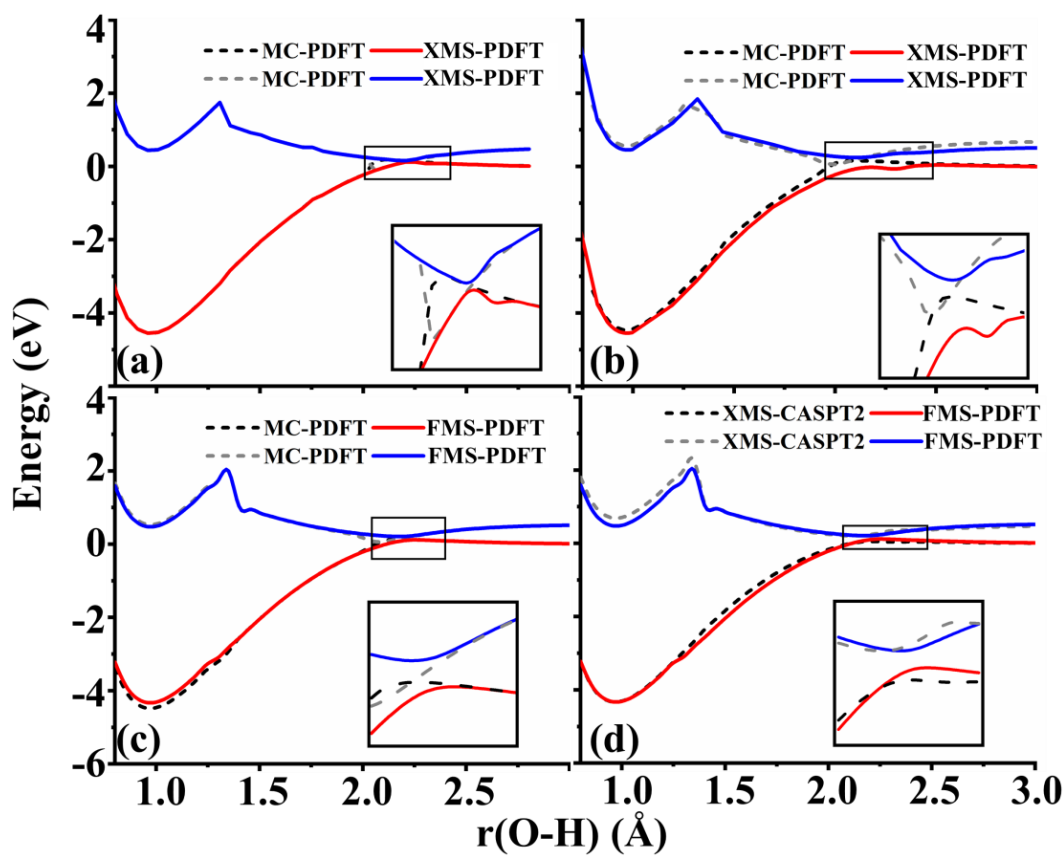

Fig. 5. (a), (b) potential energy curves of two states for O-H dissociation in phenol with the H-O-C-C dihedral angles being 1° and 10°, calculated by MC-PDFT (dash and dotted lines) and XMS-PDFT (solid lines). (c), (d) Potential energy curves for two states with a H-O-C-C dihedral angle of 10° calculated by FMS-PDFT and compared with MC-PDFT and XMS-CASPT2.

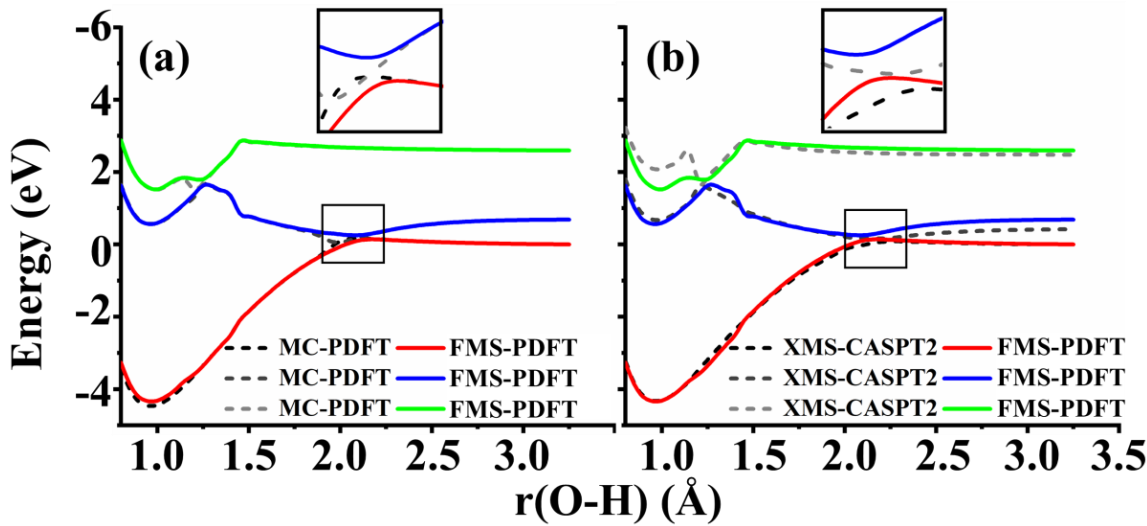

Fig. 6. Potential energy curves of three states for O-H dissociation in phenol with an H-O-C-C dihedral angle 10° as calculated by FMS-PDFT and compared to MC-PDFT and XMS-CASPT2.

Because DMRG can be used to extend CASSCF to large systems with large active spaces, we also implemented and tested FMS-PDFT based on DMRG. This test involves the two-state treatment of phenol molecule with the H-O-C-C dihedral angle at 10°. To verify the accuracy of the DRMG implementation, the active space used in FMS-PDFT/DMRG is the same as that used in FMS-PDFT. Table 5 provides the differences between FMS-PDFT/DMRG and FMS-PDFT for the rotation angles of two reference states and for the energies of the two states for a range of O-H distances. The rotation angles for generating the intermediate basis are different by no more than 0.001°, while the energies for the two states agree within 0.04 meV. The motivation for using DMRG is to study much larger active spaces with MC-PDFT (with state-specific MC-PDFT/DMRG we previously studied 30 active electrons in 30 active orbitals[26] and 34 active electrons in 35 active orbitals[27]), but the comparison presented here is to show that DMRG agrees well with a conventional solver when the conventional solver is affordable. The good agreement shows that FMS-PDFT/DMRG method is a promising method to study the PESs and dynamics of large systems.

Table 5. Differences between rotation angles ($\theta$), energies ($E_1$ and $E_2$) calculated by FMS-PDFT/DMRG and those by FMS-PDFT for each state

| $R_{O-H}$ (Å) | $\Delta\theta$ (deg) | $\Delta E_1$ (meV) | $\Delta E_2$ (meV) |
| --- | --- | --- | --- |
| 0.8 | -0.0004 | 0.007 | 0.026 |
| 1.0 | 0.0005 | -0.002 | -0.012 |
| 1.2 | 0.0001 | 0.008 | 0.007 |
| 1.4 | -0.0003 | -0.040 | -0.021 |
| 1.6 | 0.0001 | 0.016 | -0.007 |
| 1.8 | 0.0004 | 0.008 | -0.005 |
| 2.0 | 0.0000 | 0.017 | -0.005 |
| 2.2 | -0.0004 | 0.022 | 0.031 |
| 2.4 | 0.0000 | -0.007 | 0.013 |
| 2.6 | 0.0007 | -0.004 | 0.005 |
| 2.8 | 0.0000 | 0.004 | -0.006 |
| 3.0 | 0.0001 | -0.014 | 0.003 |
| 3.2 | 0.0000 | -0.006 | 0.010 |

**4.6 Oxygen atom plus oxygen molecule collision in triplet state (O + $O_2$)**

Two example cuts of the six lowest energy triplet $A'$ potential energy curves of $O_3$ system were calculated. Since three atoms are always in a plane, $C_s$ point group symmetry can be applied for this system. For the separated $O_2$ + O case, the ground energy level corresponds to the combination of an $O_2(\,^3\Sigma_g^-)$ molecule and an $O(^3P)$ atom. If only spatial degeneracy is considered, this ground energy level has three-fold degeneracy and two of the degenerate states belong to the $A'$ irrep. The first excited energy level for the separated atom and diatom corresponds to $O_2(\,^1\Delta_g)$ plus $O(^3P)$; this energy level has a six-fold spatial degeneracy, and three of these six states belong to the $A'$ irrep. Finally, the second excited energy level of the separated system corresponds to $O_2(\,^1\Sigma_g^+)$ plus $O(^3P)$ atom; this level has threefold spatial degeneracy, and one of these three states belong to the $A'$ irrep. Altogether, this makes six $^3A'$ states that are considered here (the six $^3A''$ states and the singlet and quintet states are not considered here).

The first example considered corresponds to an $O_2$ + O collision with the atom, labeled O3, impinging on the O1 end of the O1O2 diatom. The $r$(O1O2) distance is 1.208 Å, and the bond angle of the three O atoms is close to linear, α(O2O1O3) = 175°. These two geometric parameters were fixed, and the $r$(O1O3) distance, was scanned from 1.0 to 2.5 Å. Fig. 6 shows that there are several avoided crossings as O3 approaches and that the potential curves obtained by XMS-PDFT calculations agree well with those obtained by XMS-CASPT2.

Examination of the configuration interaction coefficients show that states with configurations corresponding at large $r$(O1O3) to curves $V_2$ (blue) and $V_4$ (pink), leave the six-state model space when $r$(O1O3) is decreased to ~ 2.5 Å, and two new states arrive. For $r$(O1O3) < 1.4 Å, these two new states correspond to curves $V_2$ (blue) and $V_3$ (green). It is very encouraging that XMS-PDFT agrees well with XMS-CASPT2 even for this rugged landscape with multiple avoided crossings.

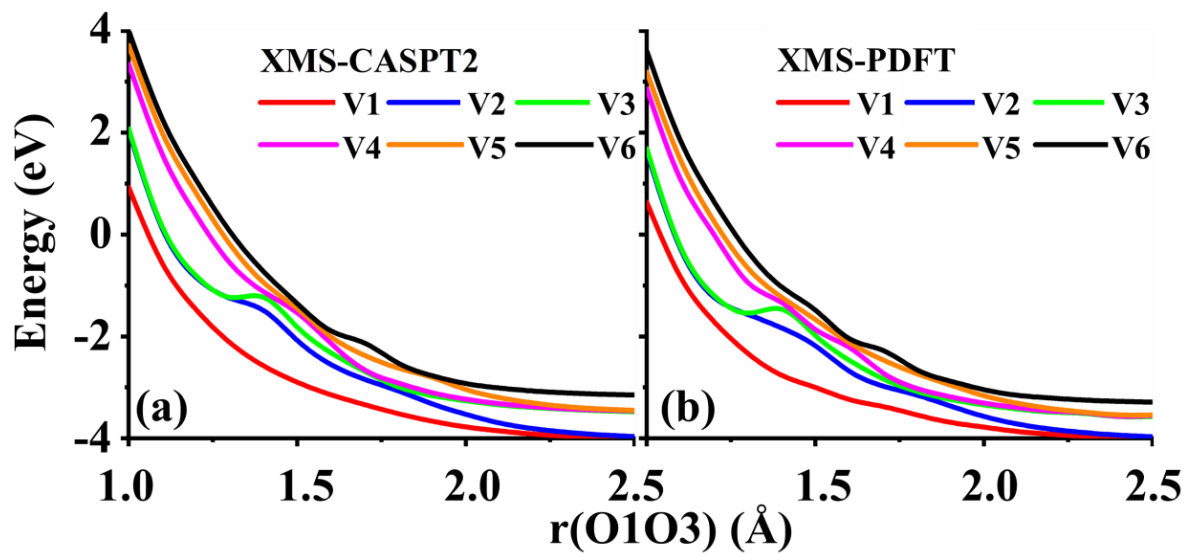

Fig. 7. Six potential energy curves of triplet O + O2 collisions calculated by XMS-CASPT2 and XMS-PDFT.

### 4.7 Triplet ozone ($O_3$)

In the second example, $r$(O1O2), is again 1.208 Å, and $r$(O1O3), is fixed at 1.4 Å. The scanning parameter is the bond angle, α(O2O1O3), varying from 60 to 180°; see Fig. 7. This example, like to the previous one, also contains several avoided crossings. However, due to the shapes of the curves, it is easier to follow the changes. The avoided crossings clearly show how the ground electronic state (corresponding to curve $V_1$) at 180° correlates to a higher energy state as the bond angle decreases. The avoided crossing of curves $V_1$ and $V_2$ is at ~160°, that of $V_2$ and $V_3$ is at ~145°, that of curves $V_3$ and $V_4$ is at ~115°, that of curves $V_4$ and $V_5$ is at ~90°, and that of curves $V_5$ and $V_6$ is at ~85°.

Again, the character of XMS-PDFT calculations are in strikingly good agreement with those obtained by XMS-CASPT2. There are, however, some minor differences, chief among which is that close to 180°, the XMS-PDFT curves are slightly more rugged than the XMS-CASPT2 curves.

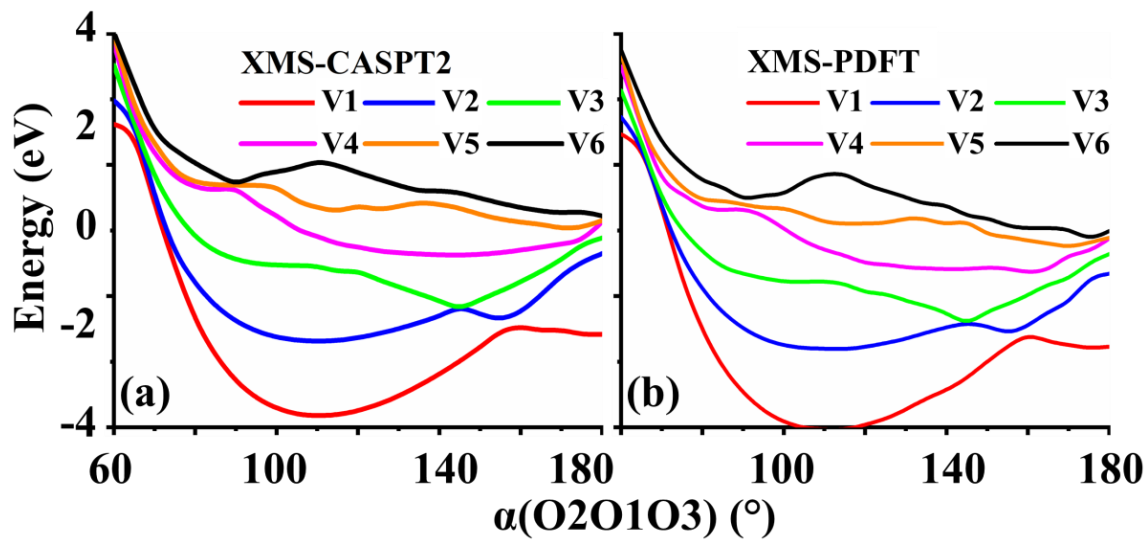

Fig. 8. Six potential energy curves of triplet ozone calculated by XMS-CASPT2 and XMS-PDFT.

### 4.8 Spiro cation

In this section, we tested 2,2′,6,6′-tetrahydro-4H,4′H-5,5′-spirobi[cyclopenta[c]pyrrole] molecule, which is simply called spiro cation in this paper. The structure of spiro cation is shown in Fig. 1 in Ref. 17. Spiro cation can be viewed as mixed-valence compound[19] that is composed of two organic subsystems (one on the left, one on the right) with a hole due to the removal of an electron to make the cation. The hole is partly localized on the left or the right subsystem, which results in their geometries being slightly different from one another. We denote the geometry when the hole is mainly on the left as geometry A, and that where the hole is mainly on the right as geometry B. Then, as in Ref. 17, we define a reaction path from geometry A to geometry B by using the linear synchronous transit method[64] as,

$$Q_\gamma(\xi) = \left(\frac{1}{2} - \xi\right) Q_\gamma^A + \left(\frac{1}{2} + \xi\right) Q_\gamma^B, \qquad \gamma = 1, 2, \dots, 3N_{\text{atoms}} \qquad (24)$$

where $Q_\gamma$ is a Cartesian coordinates for $N_{\text{atoms}}$ atoms, and $\xi$ is a parameter changing from –1.5 to 1.5. In particular, when $\xi = -0.5$ or $\xi = 0.5$, the equilibrium geometry is obtained for spiro cation and the one or the other of the adiabatic potential energy curves has a minimum. When $\xi = 0$, the geometry is average of geometries A and B, and it can be interpreted as a transition structure for intramolecular charge transfer between the left and right subsystems.

For this very difficult test case, the XMS-PDFT curves resemble the MC-PDFT curves, and the XMS-CASPT2 curves resemble the MS-CASPT2 curves, which is the expected result when the intermediate basis is the same as the CASSCF basis (zero rotation angle). The XMS-CASPT2, MC-PDFT, and FMS-PDFT potential energy curves along the path of eqn (24) are plotted in Fig. 9. It can be seen that XMS-CASPT2 and FMS-PDFT both give good results with local minima for the ground state when $\xi = \pm 0.35$ and a local maximum and avoided crossing in the ground state when $\xi = 0$. However, MC-PDFT and XMS-PDFT do not show local minima for the ground state near $\xi = \pm 0.35$ or $\pm 0.5$, and they show an unphysical dip when $\xi = 0$. The great improvement of FMS-PDFT compared with MC-PDFT again shows the value of FMS-PDFT.

Section S3 in the Supporting Information shows some other mixed-valence cases where XMS-PDFT fails to give a correct topography of PESs.

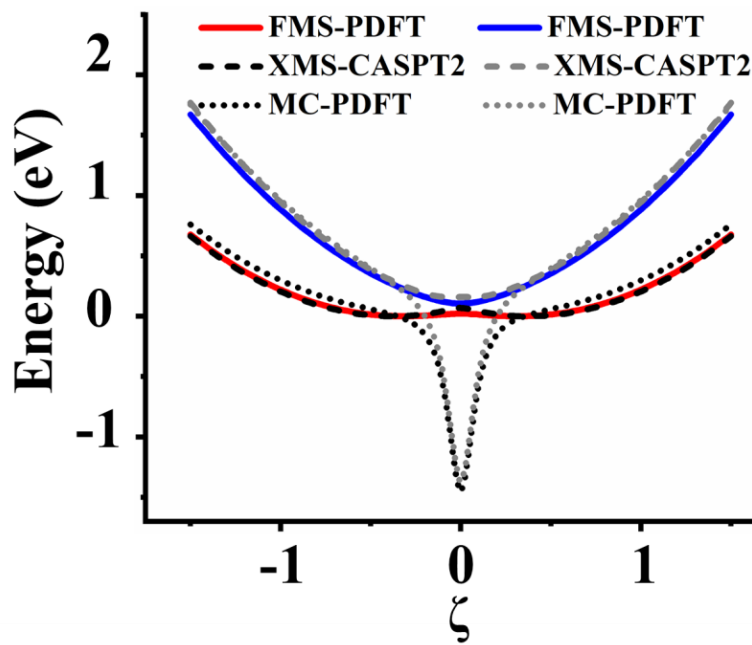

Fig. 9. Potential energy curves of two lowest states of spiro cation calculated by FMS-PDFT, XMS-CASPT2, and MC-PDFT.

## 5. Conclusions

A general scheme for multi-state MC-PDFT is proposed in this paper. In this scheme, the CASSCF reference states are rotated to a set of intermediate states via a unitary transformation, and an effective Hamiltonian matrix in the intermediate-state basis is constructed using the MC-PDFT method for the diagonal elements and wave function theory for the off-diagonal ones. Two practical methods, XMS-PDFT and VMS-PDFT, for the unitary transformations are proposed in this paper, and they are tested on eight systems exhibiting avoided crossings of two to six states. The XMS-PDFT method uses the transformation proposed by Granovsky for XMC-QDPT; VMS-PDFT chooses the transformation that maximizes the trace of the effective Hamiltonian. We implemented the VMS-PDFT method using a convenient Fourier series expansion, and the resulting method is called FMS-PDFT. Tests are performed on systems with avoided crossings to compare the two new multi-state methods, XMS-PDFT, and VMS-PDFT (mainly in the FMS-PDFT version), to state-specific MC-PDFT and the more expensive multi-state method, XMS-CASPT2. We find that FMS-PDFT, like our earlier but less convenient SI-PDFT, gives reasonable potential energy curves for all test cases examined, and it shows great improvement over MC-PDFT. Similarly XMS-PDFT gives good results for all systems except the mixed-valence spiro cation. Since XMS-PDFT is less expensive than VMS-PDFT and since it usually gives good results, we expect that both VMS-PDFT and XMS-PDFT will be useful for future work. We also implemented the FMS-PDFT method based on DMRG wave functions as a strategy proposed for calculations with large active spaces. The two new multi-state methods proposed here are preferred to the previous SI-PDFT because they treat the ground state and excited states on an equal footing and they require only a single SA-CASSCF calculation and a single set of orbitals.

**Acknowledgments**

This work was supported by the National Science Foundation under grant CHE–1746186. The work of Zoltan Varga was supported by the Air Force Office of Scientific Research by grant FA9550-19-1-0219.

**References**

1 W. Kohn, A. D. Becke and R. G. Parr, Density functional theory of electronic structure, *J. Phys. Chem.,* 1996, **100**, 12974-12980.

2 H. S. Yu, S. L. Li and D. G. Truhlar, Perspective: Kohn-Sham density functional theory descending a staircase, *J. Chem. Phys.,* 2016, **145**, 130901.

3 W. Domcke, H. Köppel and L. S. Cederbaum, Spectroscopic effects of conical intersections of molecular potential energy surfaces, *Mol. Phys.,* 1981, **43**, 851-875.

4 A. W. Jasper, B. K. Kendrick, C. A. Mead and D. G. Truhlar, Non-Born-Oppenheimer Chemistry: Potential Surfaces, Couplings, and Dynamics, in *Modern Trends in Chemical Reaction Dynamics: Experiment and Theory* (Part 1), ed. X. Yang and K. Liu, World Scientific, Singapore, 2004, pp. 329-391.

5 R. Maurice, P. Verma, J. M. Zadrozny, S. Luo, J. Borycz, J. R. Long, D. G. Truhlar and L. Gagliardi, Single-ion magnetic anisotropy and isotropic magnetic couplings in the metal–organic framework $Fe_2$(dobdc), *Inorg. Chem.,* 2013, **52**, 9379-9389.

6 M. Atanasov, J. M. Zadrozny, J. R. Long and F. Neese, A theoretical analysis of chemical bonding, vibronic coupling, and magnetic anisotropy in linear iron(II) complexes with single-molecule magnet behavior, *Chem. Sci.*, 2013, **4**, 139-156.

7 G. Li Manni, R. K. Carlson, S. Luo, D. Ma, J. Olsen, D. G. Truhlar and L. Gagliardi, Multiconfiguration pair-density functional theory, *J. Chem. Theory Comput.*, 2014, **10**, 3669-3680.

8 L. Gagliardi, D. G. Truhlar, G. Li Manni, R. K. Carlson, C. E. Hoyer and J. L. Bao, Multiconfiguration pair-density functional theory: a new way to treat strongly correlated systems, *Acc. Chem. Res.*, 2017, **50**, 66-73.

9 S. Ghosh, P. Verma, C. J. Cramer, L. Gagliardi, and D. G. Truhlar, Combining wave function methods with density functional theory for excited states, *Chem. Rev.*, 2018, **118**, 7249-7292.

10 K. Hirao, Multireference Møller-Plesset method, *Chem. Phys. Lett.*, 1992, **190**, 374-380.

11 H. Nakano, Quasidegenerate perturbation theory with multiconfigurational self-consistent-field reference functions, *J. Chem. Phys.*, 1993, **99**, 7983-7992.

12 A. A. Granovsky, Extended multi-configuration quasi-degenerate perturbation theory: the new approach to multi-state multi-reference perturbation theory, *J. Chem. Phys.*, 2011, **134**, 214113.

13 K. Andersson, P. A. Malmqvist, B. O. Roos, A. J. Sadlej and K. Wolinski, Second-order perturbation theory with a CASSCF reference function, *J. Phys. Chem.*, 1990, **94**, 5483-5488.

14 J. Finley, P.-Å. Malmqvist, B. O. Roos and L. Serrano-Andrés, The multi-state CASPT2 method, *Chem. Phys. Lett.*, 1998, **288**, 299-306.

15 T. Shiozaki, W. Győrffy, P. Celani and H.-J. Werner, Communication: extended multi-state complete active space second-order perturbation theory: energy and nuclear gradients, *J. Chem. Phys.*, 2011, **135**, 081106.

16 A. M. Sand, C. E. Hoyer, D. G. Truhlar and L. Gagliardi, State-interaction pair-density functional theory, *J. Chem. Phys.*, 2018, **149**, 024106.

17  S. S. Dong, K. B. Huang, L. Gagliardi and D. G. Truhlar, State-interaction pair-density functional theory can accurately describe a spiro mixed valence compound, *J. Phys. Chem. A*, 2019, **123**, 2100-2106.

18  C. Zhou, L. Gagliardi and D. G. Truhlar, State-interaction pair density functional theory for locally avoided crossings of potential energy surfaces in methylamine, *Phys. Chem. Chem. Phys.*, 2019, **21**, 13486-13493.

19  M. B. Robin and P. Day, Mixed Valence Chemistry: A Survey and Classification, in *Advances in Inorganic Chemistry and Radiochemistry*, eds. H. J. Emeléus and A. G. Sharpe, Academic Press 1968, vol. 10, pp. 247-422.

20  S. R. White, Density matrix formulation for quantum renormalization groups, *Phys. Rev. Lett.*, 1992, **69**, 2863-2866.

21  S. R. White, Density-matrix algorithms for quantum renormalization groups, *Phys. Rev. B*, 1993, **48**, 10345-10356.

22  K. H. Marti and M. Reiher, The density matrix renormalization group algorithm in quantum chemistry, *Z. Phys. Chem.*, 2010, **224**, 583-599.

23  Y. Kurashige and T. Yanai, Second-order perturbation theory with a density matrix renormalization group self-consistent field reference function: theory and application to the study of chromium dimer, *J. Chem. Phys.*, 2011, **135**, 094104.

24  R. Olivares-Amaya, W. Hu, N. Nakatani, S. Sharma, J. Yang and G. K.-L. Chan, The *ab-initio* density matrix renormalization group in practice, *J. Chem. Phys.*, 2015, **142**, 034102.

25  S. Knecht, E. D. Hedegård, S. Keller, A. Kovyrshin, Y. Ma, A. Muolo, C. J. Stein and M. Reiher, New approaches for *ab initio* calculations of molecules with strong electron correlation, *Chimia*, 2016, **70**, 244−251.

26  P. Sharma, V. Bernales, S. Knecht, D. G. Truhlar and L. Gagliardi, Density matrix renormalization group pair-density functional theory (DMRG-PDFT): singlet–triplet gaps in polyacenes and polyacetylenes, *Chem. Sci.*, 2019, **10**, 1716-1723.

27  C. Zhou, L. Gagliardi and D. G. Truhlar, Multiconfiguration pair-density functional theory for iron porphyrin with CAS, RAS, and DMRG active spaces, *J. Phys. Chem. A*, 2019, **123**, 3389-3394.

28  S. Wouters, W. Poelmans, P. W. Ayers and D. Van Neck, CheMPS2: A free open-source spin-adapted implementation of the density matrix renormalization group for ab initio quantum chemistry, *Comput. Phys. Commun.,* 2014, **185**, 1501-1514.

29  I. Fdez. Galván, M. Vacher, A. Alavi, C. Angeli, F. Aquilante, J. Autschbach, J. J. Bao, S. I. Bokarev, N. A. Bogdanov, R. K. Carlson, L. F. Chibotaru, J. Creutzberg, N. Dattani, M. G. Delcey, S. S. Dong, A. Dreuw, L. Freitag, L. M. Frutos, L. Gagliardi, F. Gendron, A. Giussani, L. González, G. Grell, M. Guo, C. E. Hoyer, M. Johansson, S. Keller, S. Knecht, G. Kovačević, E. Källman, G. Li Manni, M. Lundberg, Y. Ma, S. Mai, J. P. Malhado, P. Å. Malmqvist, P. Marquetand, S. A. Mewes, J. Norell, M. Olivucci, M. Oppel, Q. M. Phung, K. Pierloot, F. Plasser, M. Reiher, A. M. Sand, I. Schapiro, P. Sharma, C. J. Stein, L. K. Sørensen, D. G. Truhlar, M. Ugandi, L. Ungur, A. Valentini, S. Vancoillie, V. Veryazov, O. Weser, T. A. Wesołowski, P.-O. Widmark, S. Wouters, A. Zech, J. P. Zobel and R. Lindh, OpenMolcas: From source code to insight, *J. Chem. Theory Comput.*, 2019, **15**, 5925-5964.

30  S. Keller, M. Dolfi, M. Troyer and M. Reiher, An efficient matrix product operator representation of the quantum chemical Hamiltonian, *J. Chem. Phys.*, 2015, **143**, 244118.

31  S. Keller and M. Reiher, Spin-adapted matrix product states and operators, *J. Chem. Phys.*, 2016, **144**, 134101.
32  Y. Ma, S. Knecht, S. Keller and M. Reiher, Second-order self-consistent-field density-matrix renormalization group, *J. Chem. Theory Comput.*, 2017, **13**, 2533-2549.
33  G. Ghigo, B. O. Roos and P.-Å. Malmqvist, A modified definition of the zeroth-order Hamiltonian in multiconfigurational perturbation theory (CASPT2), *Chem. Phys. Lett.*, 2004, **396**, 142-149.
34  T. H. D. Jr., Gaussian basis sets for use in correlated molecular calculations. I. The atoms boron through neon and hydrogen, *J. Chem. Phys.*, 1989, **90**, 1007-1023.
35  E. Papajak, J. Zheng, X. Xu, H. R. Leverentz and D. G. Truhlar, Perspectives on basis sets beautiful: seasonal plantings of diffuse basis functions, *J. Chem. Theory Comput.*, 2011, **7**, 3027-3034.
36  D. Feller, The role of databases in support of computational chemistry calculations, *J. Comput. Chem.*, 1996, **17**, 1571-1586.
37  K. L. Schuchardt, B. T. Didier, T. Elsethagen, L. Sun, V. Gurumoorthi, J. Chase, J. Li and T. L. Windus, Basis set exchange: a community database for computational sciences, *J. Chem. Inf. Model.*, 2007, **47**, 1045-1052.
38  P. C. Hariharan and J. A. Pople, The influence of polarization functions on molecular orbital hydrogenation energies, *Theor. Chim. Acta*, 1973, **28**, 213-222.
39  L. R. Kahn, P. J. Hay and I. Shavitt, Theoretical study of curve crossing: *ab initio* calculations on the four lowest $^1\Sigma^+$ states of LiF, *J. Chem. Phys.*, 1974, **61**, 3530-3546.
40  B. J. Botter, J. A. Kooter and J. J. C. Mulder, Ab-initio calculations of the covalent-ionic curve crossing in LiF, *Chem. Phys. Lett.*, 1975, **33**, 532-534.
41  H. J. Werner and W. Meyer, MCSCF study of the avoided curve crossing of the two lowest $^1\Sigma^+$ states of LiF, *J. Chem. Phys.*, 1981, **74**, 5802-5807.
42  J. P. Finley and H. A. Witek, Diagrammatic complete active space perturbation theory: calculations on benzene, $N_2$, and LiF, *J. Chem. Phys.*, 2000, **112**, 3958-3963.
43  J. Meller, J.-P. Malrieu and J.-L. Heully, Size-consistent multireference configuration interaction method through the dressing of the norm of determinants, *Mol. Phys.*, 2003, **101**, 2029-2041.
44  Ö. Legeza, J. RÖDer and B. A. Hess, QC-DMRG study of the ionic-neutral curve crossing of LiF, *Mol. Phys.*, 2003, **101**, 2019-2028.
45  C. Angeli, R. Cimiraglia and J.-P. Malrieu, A simple approximate perturbation approach to quasi-degenerate systems, *Theor. Chem. Acc.*, 2006, **116**, 434-439.
46  M. Hanrath, Multi-reference coupled-cluster study of the ionic-neutral curve crossing LiF, *Mol. Phys.*, 2008, **106**, 1949-1957.
47  F. Sasaki and M. Yoshimine, Configuration-interaction study of atoms. II. electron affinities of B, C, N, O, and F, *Phys. Rev. A*, 1974, **9**, 26-34.
48  D. G. Truhlar and C. A. Mead, Relative likelihood of encountering conical intersections and avoided intersections on the potential energy surfaces of polyatomic molecules, *Phys. Rev. A*, 2003, **68**, 32501.

49 H. Nakamura and D. G. Truhlar, Extension of the fourfold way for calculation of global diabatic potential energy surfaces of complex, multiarrangement, non-Born-Oppenheimer systems: application to HNCO($S_0$,$S_1$), *J. Chem. Phys.*, 2003, **118**, 6816-6829.

50 J. V. Michael and W. A. Noyes, The photochemistry of methylamine, *J. Am. Chem. Soc.*, 1963, **85**, 1228-1233.

51 E. Kassab, J. Gleghorn and E. Evleth, Theoretical aspects of the photochemistry of methanol, methylamine, and related materials, *J. Am. Chem. Soc.*, 1983, **105**, 1746-1753.

52 G. Waschewsky, D. Kitchen, P. Browning and L. Butler, Competing bond fission and molecular elimination channels in the photodissociation of $CH_3NH_2$ at 222 nm, *J. Phys. Chem.*, 1995, **99**, 2635-2645.

53 C. L. Reed, M. Kono and M. N. R. Ashfold, Near-UV photolysis of methylamine studied by H-atom photofragment translational spectroscopy, *J. Chem. Soc., Faraday Trans.*, 1996, **92**, 4897-4904.

54 K. M. Dunn and K. Morokuma, Ab initio study of the photochemical dissociation of methylamine, *J. Phys. Chem.*, 1996, **100**, 123-129.

55 S. J. Baek, K.-W. Choi, Y. S. Choi and S. K. Kim, Spectroscopy and dynamics of methylamine. I. Rotational and vibrational structures of $CH_3NH_2$ and $CH_3ND_2$ in Ã states, *J. Chem. Phys.*, 2003, **118**, 11026-11039.

56 M. H. Park, K.-W. Choi, S. Choi, S. K. Kim and Y. S. Choi, Vibrational structures of methylamine isotopomers in the predissociative Ã states: $CH_3NHD$, $CD_3NH_2$, $CD_3NHD$, and $CD_3ND_2$, *J. Chem. Phys.*, 2006, **125**, 084311.

57 D.-S. Ahn, J. Lee, J.-M. Choi, K.-S. Lee, S. J. Baek, K. Lee, K.-K. Baeck and S. K. Kim, State-selective predissociation dynamics of methylamines: The vibronic and H/D effects on the conical intersection dynamics, *J. Chem. Phys.*, 2008, **128**, 224305.

58 C. Levi, R. Kosloff, Y. Zeiri and I. Bar, Time-dependent quantum wave-packet description of H and D atom tunneling in N–H and N–D photodissociation of methylamine and methylamine-d2, *J. Chem. Phys.*, 2009, **131**, 064302.

59 R. Marom, C. Levi, T. Weiss, S. Rosenwaks, Y. Zeiri, R. Kosloff and I. Bar, Quantum tunneling of hydrogen atom in dissociation of photoexcited methylamine, *J. Phys. Chem. A*, 2010, **114**, 9623-9627.

60 D.-S. Ahn, J. Lee, Y. C. Park, Y. S. Lee and S. K. Kim, Nuclear motion captured by the slow electron velocity imaging technique in the tunnelling predissociation of the $S_1$ methylamine, *J. Chem. Phys.*, 2012, **136**, 024306.

61 J. O. Thomas, K. E. Lower and C. Murray, Observation of NH $X^3\Sigma^-$ as a primary product of methylamine photodissociation: Evidence of roaming-mediated intersystem crossing? *J. Phys. Chem. Lett.*, 2012, **3**, 1341-1345.

62 H. Xiao, S. Maeda and K. Morokuma, Theoretical study on the photodissociation of methylamine involving $S_1$, $T_1$, and $S_0$ states, *J. Phys. Chem. A*, 2013, **117**, 5757-5764.

63 M. Epshtein, Y. Yifrach, A. Portnov and I. Bar, Control of nonadiabatic passage through a conical intersection by a dynamic resonance, *J. Phys. Chem. Lett.*, 2016, **7**, 1717-1724.

64 T. A. Halgren and W. N. Lipscomb, The synchronous-transit method for determining reaction pathways and locating molecular transition states, *Chem. Phys. Lett.*, 1977, **49**, 225-232.